\documentclass[a4paper,11pt]{article}
\pdfoutput=1 

\usepackage{jinstpub} 

\title{\boldmath First results from the upgrade of the Extreme Energy Events experiment}

\author[1,2,*]{M.~Abbrescia,\note{Corresponding author, email: marcello.abbrescia@ba.infn.it}}
\author[1,3]{C.~Avanzini,}
\author[1,3]{L.~Baldini,}
\author[1,4]{R.~Baldini Ferroli,}
\author[1,3]{G.~Batignani,}
\author[1,17]{M.~Battaglieri,}
\author[1,8]{S.~Boi,}
\author[1,5]{E.~Bossini,}
\author[1,6]{F.~Carnesecchi,}
\author[1,7]{A.~Chiavassa,}
\author[1,8]{C.~Cicalo,}
\author[1,6]{L.~Cifarelli,}
\author[1]{F.~Coccetti,}
\author[1,9]{E.~Coccia,}
\author[1,10]{A.~Corvaglia,}
\author[1,11]{D.~De~Gruttola,}
\author[1,11]{S.~De Pasquale,}
\author[1,4]{L.~Fabbri,}
\author[16]{ V.~Frolov,}
\author[1,7]{L.~Galante,}
\author[1,7]{P.~Galeotti,}
\author[1,6]{M.~Garbini,}
\author[1,17]{G.~Gemme,}
\author[1,7]{I.~Gnesi,}
\author[1]{ S.~Grazzi,}
\author[1,12]{C.~Gustavino,}
\author[1,6,15]{D.~Hatzifotiadou,}
\author[1,18]{P.~La~Rocca,}
\author[1,19]{G.~Mandaglio,}
\author[14]{O.~Maragoto Rodriguez,}
\author[13]{G.~Maron,}
\author[1,20]{M.~N.~Mazziotta,}
\author[1,4]{S.~Miozzi,}
\author[1,6]{R.~Nania,}
\author[1,6]{F.~Noferini,}
\author[1,21]{F.~Nozzoli,}
\author[1,6]{F.~Palmonari,}
\author[1,10]{M.~Panareo,}
\author[1,10]{M.~P.~Panetta,}
\author[1,5]{R.~Paoletti,}
\author[14]{W.~Park,}
\author[1,6]{C.~Pellegrino,}
\author[1,17]{L.~Perasso,}
\author[1,3]{F.~Pilo,}
\author[1,7]{G.~Piragino,}
\author[1,4]{S.~Pisano,}
\author[1,18]{F.~Riggi,}
\author[1]{G.~C.~Righini,}
\author[1,11]{C.~Ripoli,}
\author[1,2]{M.~Rizzi,}
\author[1,6]{G.~Sartorelli,}
\author[1,6]{E.~Scapparone,}
\author[1,22]{M.~Schioppa,}
\author[1,3]{A.~Scribano,}
\author[1,6]{M.~Selvi,}
\author[1,8]{S.~Serci,}
\author[1,17]{S.~Squarcia,}
\author[1,17]{M.~Taiuti,}
\author[1,3]{G.~Terreni,}
\author[1,23]{A.~Trifir\`{o},}
\author[1,23]{M.~Trimarchi,}
\author[13]{M.~C.~Vistoli,}
\author[1,12]{L.~Votano,}
\author[1,6,15]{M.~C.~S.~Williams,}
\author[1,14,15]{L.~Zheng,}
\author[1,6,15]{A.~Zichichi,}
\author[1,15]{R.~Zuyeuski,}

\affiliation[1]{Museo Storico della Fisica e Centro Studi e Ricerche Enrico Fermi, Roma, Italy}
\affiliation[2]{INFN and Dipartimento Interateneo di Fisica, Universit\`{a}
	di Bari, Bari, Italy}
\affiliation[3]{INFN and Dipartimento di Fisica, Universit\`{a} di Pisa,
	Pisa, Italy}
\affiliation[4]{INFN, Laboratori Nazionali di Frascati, Frascati (RM),
	Italy}
\affiliation[4]{INFN, Laboratori Nazionali di Frascati, Frascati (RM),
	Italy}
\affiliation[5]{INFN Gruppo Collegato di Siena and Dipartimento di Fisica,
	Universit\`{a} di Siena, Siena, Italy}
\affiliation[6]{INFN and Dipartimento di Fisica e Astronomia,
	Universit\`{a} di Bologna, Bologna, Italy}
\affiliation[7]{INFN and Dipartimento di Fisica, Universit\`{a} di Torino,
	Torino, Italy}
\affiliation[8]{INFN and Dipartimento di Fisica, Universit\`{a} di
	Cagliari, Cagliari, Italy}
\affiliation[9]{INFN and Dipartimento di Fisica, Universit\`{a} di Roma Tor
	Vergata, Roma, Italy}
\affiliation[10]{INFN and Dipartimento di Matematica e Fisica,
	Universit\`{a} del Salento, Lecce, Italy}
\affiliation[11]{INFN and Dipartimento di Fisica, Universit\`{a} di
	Salerno, Salerno, Italy}
\affiliation[12]{INFN, Laboratori Nazionali del Gran Sasso, Assergi (AQ),
	Italy}
\affiliation[13]{INFN CNAF, Bologna, Italy}
\affiliation[14]{ICSC World Laboratory, Geneva, Switzerland}
\affiliation[15]{CERN, Geneva, Switzerland}
\affiliation[16]{JINR Joint Institute for Nuclear Research, Dubna, Russia}
\affiliation[17]{INFN and Dipartimento di Fisica, Universit\`{a} di Genova,
	Genova, Italy}
\affiliation[18]{INFN and Dipartimento di Fisica e Astronomia,
	Universit\`{a} di Catania, Catania, Italy}
\affiliation[19]{INFN Sezione di Catania and Dipartimento di Scienze
	Chimiche, Biologiche, Farmaceutiche e Ambientali, Universit\`{a} di
	Messina, Messina, Italy}
\affiliation[20]{INFN Sezione di Bari, Bari, Italy}
\affiliation[21]{Trento Institute for Fundamental Physics and Applications,Trento, Italy}
\affiliation[22]{INFN and Dipartimento di Fisica, Universit\`{a} della
	Calabria, Cosenza, Italy}
\affiliation[23]{INFN Sezione di Catania and Dipartimento di Scienze Matematiche e Informatiche, Scienze Fisiche e Scienze della Terra, Universit\'a di Messina, Messina, Italy}

\emailAdd{collaborationEEE@centrofermi.it}
\abstract{The Extreme Energy Events (EEE) experiment is the largest system in the world completely implemented with Multigap Resistive Plate Chambers (MRPCs). Presently, it consists of a network of 59 muon telescopes, each made of 3 MRPCs, devoted to the study of secondary cosmic rays. Its stations, sometimes hundreds of kilometers apart, are synchronized at a few nanoseconds level via a clock signal delivered by the Global Positioning System. The data collected during centrally coordinated runs are sent to INFN CNAF, the largest center for scientific computing in Italy, where they are reconstructed and made available for analysis. Thanks to the on-line monitoring and data transmission, EEE operates as a single coordinated system spread over an area of about $3 \times 10^5$ km$^2$.

In 2017, the EEE collaboration started an important upgrade program, aiming to extend the network with 20 additional stations, with the option to have more in the future. This implies the construction, testing and commissioning of 60 chambers, for a total detector surface of around 80 m$^2$. In this paper, aspects related to this challenging endeavor are covered, starting from the technological solutions chosen to build these state-of-the-art detectors, to the quality controls and the performance tests carried on.
}

\keywords{Resistive-plate chambers, Gaseous detectors}

\arxivnumber{2292053} 
\collaboration{The EEE collaboration}
\proceeding{The XIV$^{\text{th}}$ Workshop on Resistive Plate Chambers and related detectors\\
Feb. 19-23/2018 \\
Puerto Vallarta, Jalisco State, MEXICO}
\begin{document}
\maketitle
\flushbottom

\section{Introduction}
\label{sec:intro}

The Extreme Energy Events (EEE) project is, at the moment, the most extensive experiment to detect secondary cosmic particles in Europe, consisting in more than 50 telescopes distributed all over the Italian territory, plus two at CERN. It is funded and coordinated by the "Enrico Fermi" Historical Museum of Physics and Study and Research Centre, in collaboration with the Italian National Institute for Nuclear Physics (INFN), CERN, and several Italian universities \cite{CentroFermi}. 

Each EEE telescope comprises three layers of  Multi-gap Resistive Plate Chambers (MRPCs in the following), made in glass, 80 $\times 160$ cm$^2$ in dimensions, and equipped with six 300 $\mu$m gas gaps each, built using the same technology developed for the ALICE time-of-flight detector at LHC \cite{AliceTOF}. The EEE telescopes are capable of reconstructing the trajectories of the charged particles traversing them with an about half degree angular precision, and a time resolution at the chamber level of, roughly, two hundreds picoseconds \cite{LargeArea} \cite{Performance} \cite{DeGruttola}. The raw data from all telescopes of the network are conveyed at the largest center for scientific computing in Italy, namely CNAF in Bologna, managed by INFN, where they are automatically reconstructed and stored for subsequent analysis.

The excellent performance of the EEE telescopes allows a large variety of studies, ranging from the measure of the local muon flux \cite{Status}, the observation of variations in time of the rate of cosmic muons correlated with astrophysical events (usually called "Forbush" decreases) \cite{Forbush}, the search for anisotropies in the muon angular distribution \cite{Looking}, to the detection of extensive air showers producing time correlations in EEE stations in the same metropolitan area \cite{Showers}, or the search for large-scale correlations between showers detected in telescopes tens, hundreds, or thousands of kilometers apart \cite{Long}.

During the last years the experiment entered its maturity phase, with a great boost being given by organizing simultaneous and centralized data taking, with the whole array of telescopes taking data at the same time. This allowed to accumulate a huge statistics of candidate muon tracks, totaling now around 70 billion events. 

In 2017, the EEE collaboration decided to start an important upgrade program, aiming to extend the network with, at least, 20 additional stations. This implied the construction of sixty new chambers, for a total surface of new detectors to be built of around 80 m$^2$, comparable to the extensions foreseen for the RPC upgrade programs of some LHC experiments. Moreover, the new chambers present characteristics slightly different with respect to the ones already in use in the EEE network, to improve their performance in operation.

In order to assure an optimal performance of the new chambers, a thorough procedure of quality control has been developed, consisting in a series of tests devoted to check that these chambers satisfy criteria on electrical connectivity, gas tightness, current and efficiency. The characteristics of the new chambers and the tests performed are discussed in more details in the following. 

\section{Improvements to the EEE system for the upgrade}
\label{sec:description}

The chambers for the EEE upgrade present some changes with respects to the ones already installed in the EEE network and presently taking data. In particular, their gas gap thickness has been reduced, from 300 to 250 $\mu$m. The reason behind this decision is related to the fact that EEE is planning to operate soon its stations with new eco-friendly gas mixtures, whose study is currently going on (see, for details, \cite{Pisano}). Being based on tetrafluoropropane, these are generally characterized by a higher operating voltage with respect to the standard gas mixtures, made up of tetrafluoroethane and sulphur hexafluoride, used in EEE. Reducing the gas gap thickness will have the advantage to limit the total voltage applied to the multigap structure; this, in the case of the EEE chambers, should not exceed 20 kV, mainly because of technological reasons related to the HV connectors and DC-DC voltage converters used to feed the chambers. With 250 $\mu$m gaps, already some encouraging results have been obtained, with operating voltage and efficiency in the correct range \cite{Pisano}. To distinguish the new chambers, a green band is painted on the metallic frame enclosing the multigap structure.

Also the production of a new bunch of front-end boards has been required, due to the shortage of old ones. The EEE collaboration has taken the occasion of slightly re-designing these boards, changing the Amphenol connectors and cables currently used, with Nugent-Robinson ones, less expensive and easier to use. 

In addition, the EEE telescopes data acquisition system has been improved. In particular, a new trigger/GPS combo card was designed, which joins in the same board the functionalities of the trigger card currently used in the EEE telescopes with the ones provided by the commercial GPS modules used to synchronize the various EEE stations scattered over Italy. The new boards also provides other functionalities, such us a remotely programmable trigger logic, counters to monitor the chambers counting rates, and the distribution of a precise clock to the two TDC modules used for EEE chambers readout, necessary to synchronize them \cite{Panetta}.

In 2017, twenty-seven new chambers have been produced, twenty-four of them sufficient to equip eight new EEE stations, while the rest are kept as spares in case replacements would be needed. Most of them have already been shipped to their final destinations, and some of the corresponding stations have already started taking data, while others are in various commissioning phases. Moreover, in the first three months of 2018, other five chambers have been produced. The results reported later on in this paper refer to the total set of 33 new chambers produced up to now.

\section{Test procedure for quality control, and results}
\label{sec:tests}

The tests performed on the chambers built for the upgrade of the EEE experiment fall in three categories, executed in sequence, and described in the following.

\subsection{Connectivity checks}

These checks are performed during chamber construction, and their goal is to assure a good quality of the high voltage and readout connections. In particular, one of the issues sometimes encountered in the past was related to internal cabling, and thus a setup to perform an electrical test of all the readout strips automatically, without the need for a signal generator and oscilloscope, was developed \cite{Bossini}.



\begin{figure}[tbp]
\centering 
\includegraphics[width=1.0\textwidth]{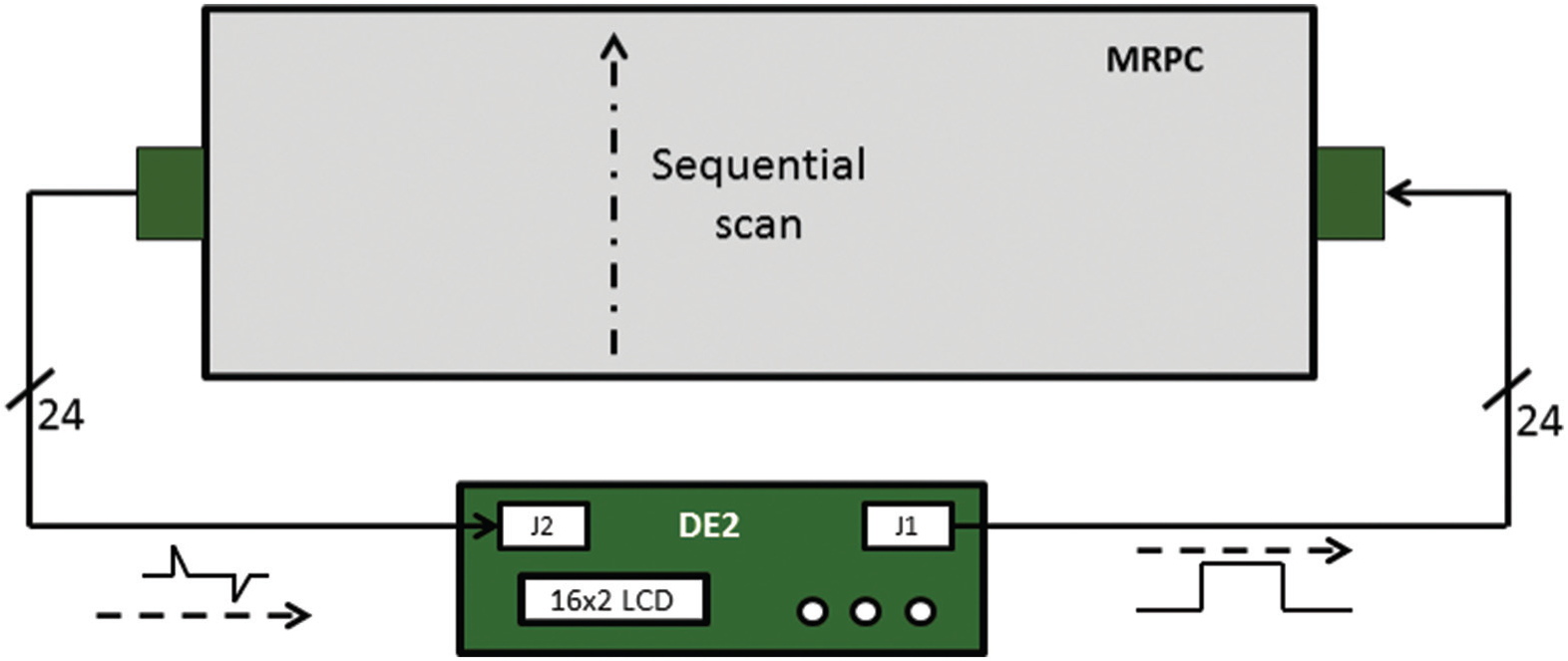}
\caption{\label{fig:Setup} Scheme of the setup used for the electric test.}
\end{figure}

The setup is based on a Terasic development board, hosting an Altera FPGA
Cyclon II with a custom firmware and an LCD screen. The 2 on-board expansion
connectors are connected to the chamber interface card on each side (see Figure~\ref{fig:Setup}) by means of passive adapters. A square digital signal is generated inside the
FPGA and sequentially pulsed on each strip. The signal is expected to
travel along the strip and come back to the FPGA through the connector on the opposite side of chamber. An automatic scan is stared, cycling on all strips, and if any issue is found the system generates an error list that contains strip by strip error information. Applied to the chambers built for the upgrade, the test was able to disentangle and identify any problem related to cabling, before the chambers were sealed and connected to the acquisition chain.
 
\subsection{Gas tightness tests}

\begin{figure}[btp]
\centering 
\includegraphics[width=1.0\textwidth]{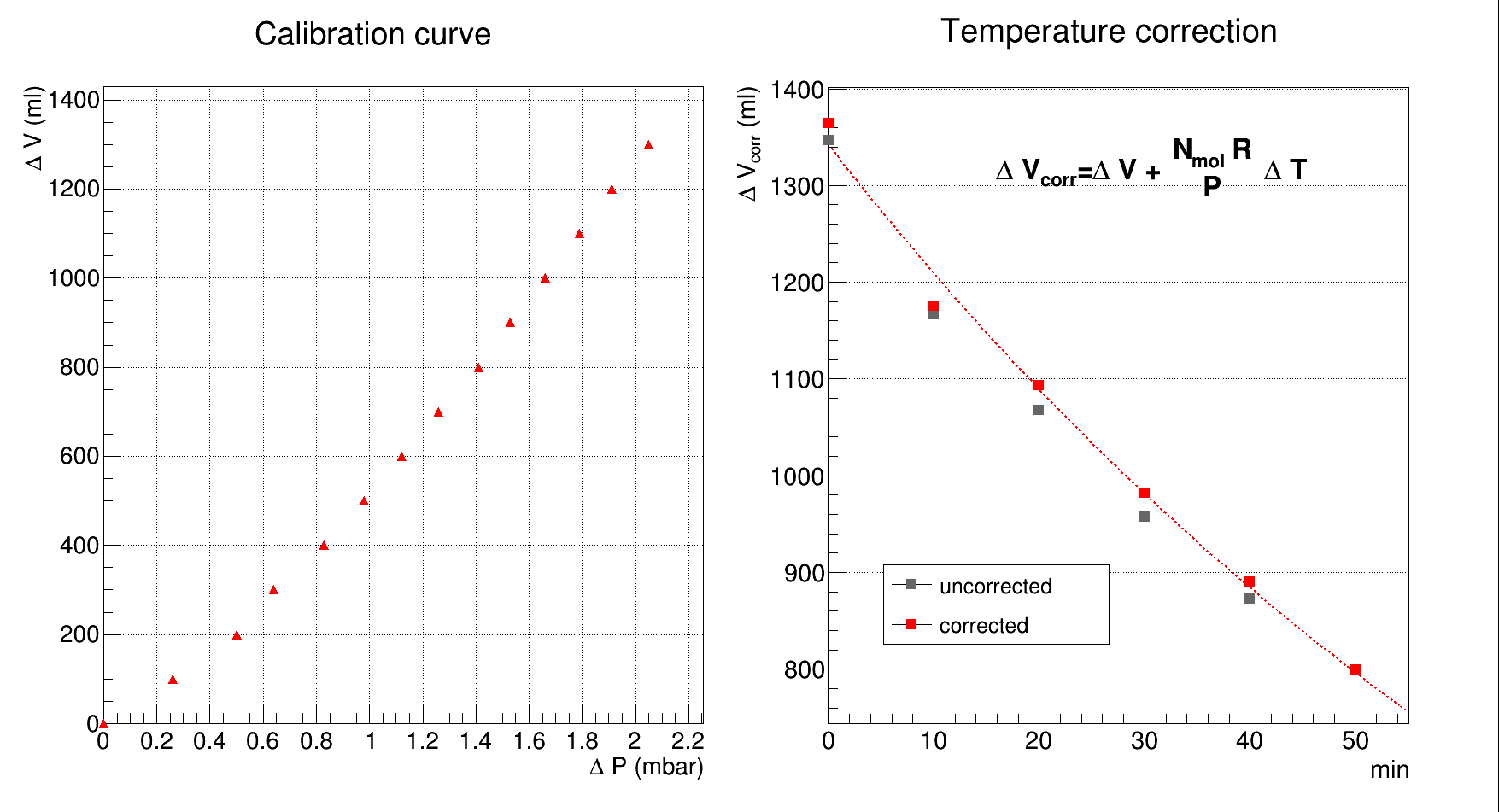}
\caption{\label{fig:Leak} Left: Calibration curve, namely the overpressure $\Delta p$ measured inside a chamber vs. the volume $\Delta V$ of air injected inside it. Right: Volume of air still remaining in the chamber vs. time, measured during a leak test; two sets of points are shown: the ones directly deduced from the calibration plots (labeled as "uncorrected") and the values for $\Delta V_{corr}$ computed applying the temperature correction reported in the formula.}
\end{figure}

These tests are devoted to check that the metallic frames enclosing the detectors have been assembled in such a way to assure that the gas mixture inside does not leak in a significant amount to the exterior. During these tests, a chamber is at first injected with air, in steps of 100 ml each, and the corresponding overpressure $\Delta p$ with respect to the atmosphere is measured. Air injection continues until a 2 mbar overpressure is reached, typical of the operation of these chambers at the stations of the EEE experiment.
One of the plots obtained during this operation, conventionally called "calibration plots", is shown in Figure~\ref{fig:Leak} (Left).

Then the chamber inlet is closed, and overpressure vs. time is measured, during an hour time span. The measured overpressure is converted, by means of the calibration curve previously obtained, to estimate the gas volume $\Delta V$ injected, and still remaining inside the chamber. To improve the precision of the measurements, corrections are applied to take into account possible temperature changes during the measure, and a "corrected" $\Delta V_{corr}$ is computed by applying the formula: 

\begin{equation}
\Delta V_{corr} = \Delta V + \frac{N_{mol} R}{p} \Delta T
\end{equation}

\noindent where $\Delta V$ is the volume of air still inside the chamber as inferred by the calibration plot, $N_{mol}$ is the total number of moles of the gas inside the chamber, $R$ is the constant of ideal gases, and $\Delta T$ is the difference in temperature between the moment the chamber inlet is closed, and the times when the experimental points are taken during the leak test. An exponential fit is used to interpolate the experimental points, and the volume leak rate can be inferred from the derivative of the fitting curve.
An example of the results obtained applying this procedure is shown in Figure~\ref{fig:Leak} (Right).

\begin{figure}[tbp]
\centering 
\includegraphics[width=0.6\textwidth]{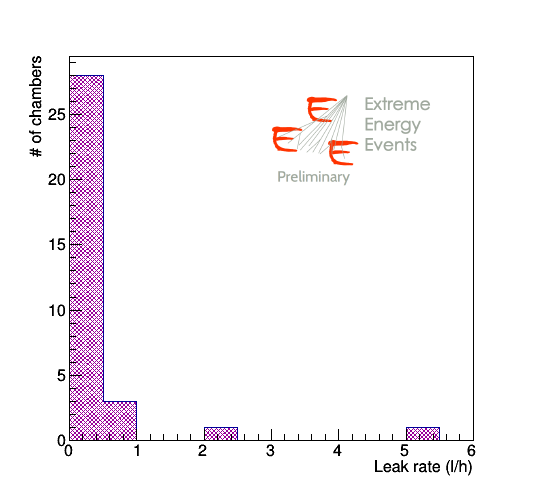}
\caption{\label{fig:LeakStatistics} Distribution of the leak rates measured for the 33 chambers of the first upgrade bunch.}
\end{figure}

Chambers are accepted if their leak rate is less than one liter per hour, condition that has been met by 31 chambers out of 33 of the set of chambers considered here; the average leak rate for the accepted chambers was 0.17 l/h (see Figure~\ref{fig:LeakStatistics}). Two chambers, characterized by a higher leak rate, were subsequently opened and the metallic frame checked to find critical spots, and then closed again. 

\subsection{Tests on current, counting rate and efficiency}

After the gas tightness tests, chambers are moved and put on top of one of the telescopes of the EEE network, namely CERN-01, where they are fluxed with the same gas mixture flowing in CERN-01 (C$_2$H$_2$F$_4$/SF$_6$ 98/2), for about four days. 

CERN-01 is used as a trigger and tracking system, in order to select cosmic muons also traversing the chambers above it. Chambers are tested one by one, reading their data by connecting their front-end boards to multi-hit TDC modules, equal to the ones used for the readout of the standard EEE telescope chambers, and triggered by the trigger signals provided by CERN-01. Also the DAQ program used is the same.

Chambers are switched on, and a 21 kV operating voltage is applied for 12-14 hours. This period, called "conditioning" was limited in time, due the constraints related to the coordinated data taking with the rest of the EEE network, or due to the construction and testing of other bunches of chambers. 

Current, counting rate and efficiency are measured by varying the applied voltage in 0.5 kV steps. Current is measured directly on the DC-DC converters providing the high voltage to the chamber under test. 
Counting rate is measured using the OR from the 24 strips equipping each chamber provided by the front-end boards, and feeding them to suitable counters. Efficiency is measured by searching hits on the chamber under test in a fiducial zone of $\pm$ 2 strips around the track extrapolated from CERN-01, and counting the number of times a hit is found in the zone and dividing it by the total number of tracks extrapolated.

\begin{figure}[tbp]
\centering 
\includegraphics[width=1.0\textwidth]{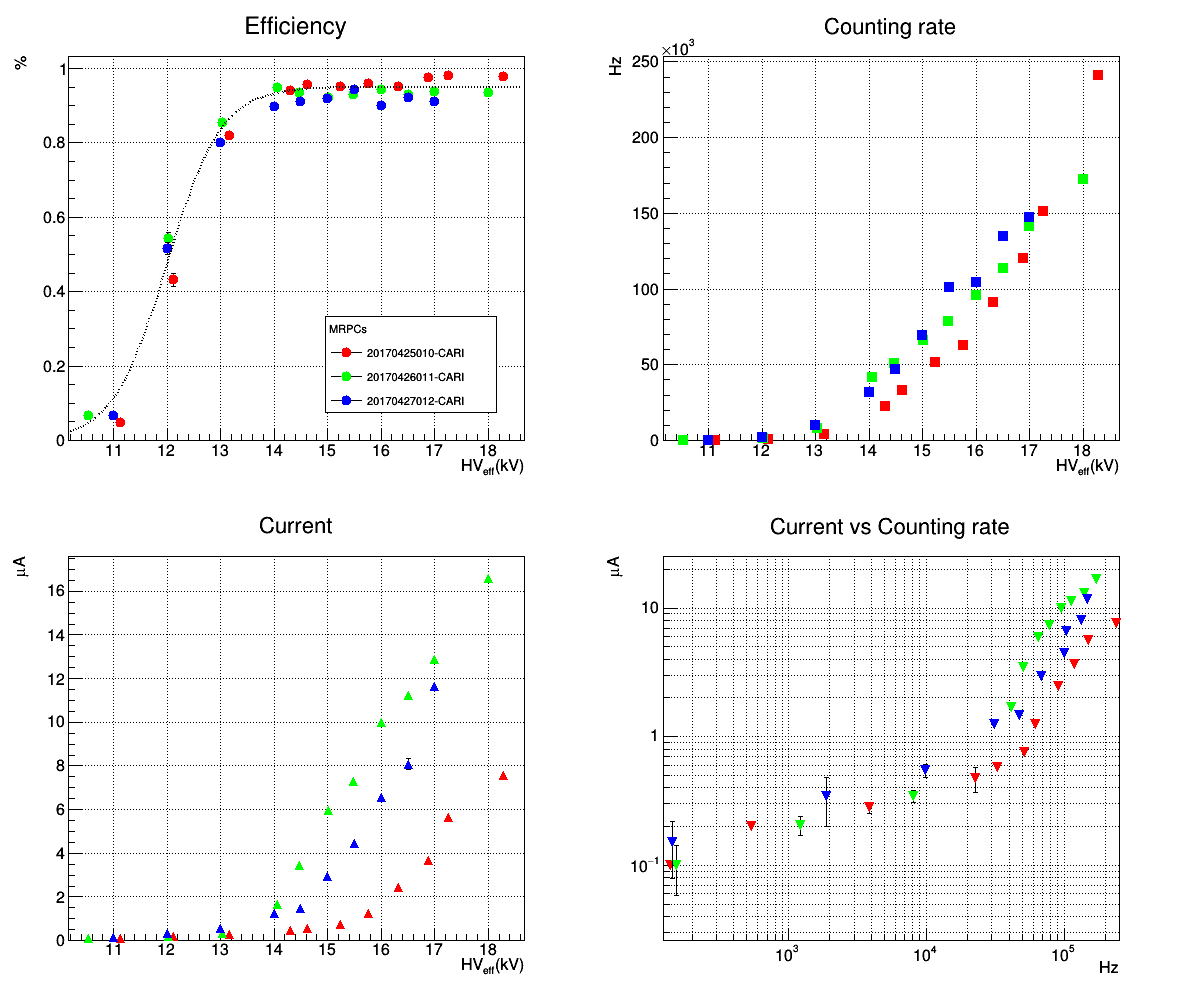}
\caption{\label{fig:PerformanceCariati} Some plots obtained on the set of chambers subsequently sent to Cariati, one of the stations of the EEE network. For the three chambers of this test, are reported: Top left: Efficiency vs. effective operating voltage $HV_{eff}$; the sigmoid superimposed is just to guide the eye; Top right: Counting rate vs. $HV_{eff}$; Bottom left: Current vs. $HV_{eff}$; Bottom right: Current vs. counting rate.}
\end{figure}

Examples of results, for a set of chambers now installed in the EEE telescope at Cariati, are shown in Figure~\ref{fig:PerformanceCariati}. Note that the applied operating voltage has been converted to an "effective" operating voltage, using the formula: 

\begin{equation}
HV_{eff} = HV_{app} \frac{T}{T_0}\frac{p_0}{p}
\end{equation}

\noindent where $p$ and $T$ are the actual values of temperature and pressure, $p_0$ and $T_0$ are some reference values for temperature and pressure arbitrarily chosen. This is a well known procedure used to compare results obtained in conditions of temperature and pressure changing with time \cite{Temperature}. In this case $p_0$ = 1000 mbar and $T_0$ = 20 \textdegree C.

\begin{figure}[tbp]
\centering 
\includegraphics[width=0.49\textwidth]{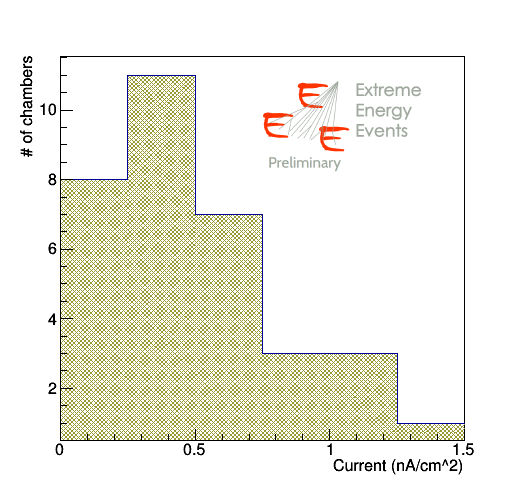}
\includegraphics[width=0.49\textwidth]{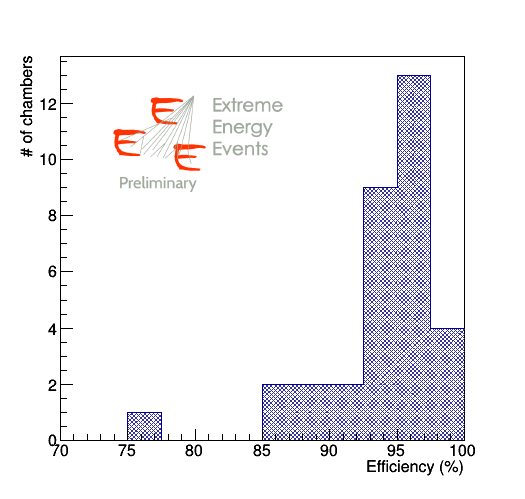}
\caption{\label{fig:PerformanceSummary} Left: Current, measured at 18 kV, for the 33 chambers manufactured up to now for the upgrade of the EEE experiment; (Right): Plateau efficiency for the same set of chambers}
\end{figure}

Some overall statistics, obtained on the whole set of 33 chambers considered here, are reported in Figure~\ref{fig:PerformanceSummary}. In this case, current, shown in Figure~\ref{fig:PerformanceSummary} (Left), has been measured at 18 kV. Given the short conditioning time, all chambers with a current lower than 2 nA/cm$^2$ are accepted. 
We have experienced that the current and the counting rate of these chambers significantly decrease, of one or even two orders of magnitude, after they have been kept operating for a few days at their nominal operating voltage.

Experimental values for efficiency vs. operating voltage are fitted using a sigmoid curve; such a fit has been performed for each chamber, and the parameter corresponding to the plateau efficiency was used to fill the histogram reported in Figure~\ref{fig:PerformanceSummary} (Right). Chamber efficiency distribution is centered around 95\%, quite an acceptable value considering that it has been obtained with cosmic rays on recently produced chambers, conditioned for a limited time. Chambers are accepted if their plateau efficiency is higher than 85\%, requirement met by 32 chambers out of 33; the chamber characterized by a plateau efficiency around 78\% is being kept under observation to understand the reason of such lower efficiency.

Not to lose any information, and to keep track of the chamber performance for possible comparison with their behavior during data taking in the coming years, everything is stored in a dedicated upgrade construction database.

\section{Conclusions}
\label{sec:conclusions}

The upgrade program of the EEE experiment is an endeavor that is currently employing the EEE collaboration, and will last until the end of 2019. It is of paramount importance to assure that the new chambers produced satisfy strict requirements on their performance, since, once they are deployed in operation, additional interventions are much more difficult to perform than when they are freshly produced at CERN. 

The tests performed have verified that the construction procedures followed assure a quality of chambers fully compatible with the requirements of the EEE experiment. In the first bunch of chambers produced, two needed to be re-opened to fix gas leak problems, one had to be kept under observation and its efficiency measured again. All the others could be immediately shipped to their final destinations and assembled in telescopes. Some of these telescopes have already started taking data.  

The additional stations available after the upgrade will improve the capabilities of the EEE network to carry on its research program. The statistics of candidate muon tracks will be rapidly increased by the new stations progressively entering into operation, and this will bring benefit to the whole spectrum of studies performed by the EEE collaboration.
Particular benefit will arrive to the searches for rare phenomena, like possible long distance correlations, which the EEE experiment is in an almost unique condition to pursue. 


%
%

\end{document}